\documentclass[11pt,english]{article}
\usepackage[T1]{fontenc}
\usepackage[utf8]{luainputenc}
\usepackage{mathrsfs}
\usepackage{amstext}
\usepackage{amssymb}
\usepackage{setspace}
\usepackage{esint}
\makeatletter
\newcommand{\lyxaddress}[1]{
\par {\raggedright #1
\vspace{1.4em}
\noindent\par}
}

\usepackage{babel}

\makeatother

\usepackage{babel}
\begin{document}

\title{Derivation of Hamilton-like equations on a non-Cauchy hypersurface
and their expected connection to quantum gravity theories}

\author{Merav Hadad , Levy Rosenblum \\
{\small{}\hspace{0.2in} }}
\maketitle

\lyxaddress{{\small{}{}Department of Natural Sciences, The Open University of
Israel P.O.B. 808, Raanana 43107, Israel. E-mail: meravha@openu.ac.il
; levyros@gmail.com}{\large{}{} }}
\begin{abstract}
Recently it was found that quantum gravity theories may involve constructing
a quantum theory on non-Cauchy hypersurfaces. However this is problematic
since the ordinary Poisson brackets are not causal in this case. We
suggest a method to identify classical brackets that are causal on
2+1 non-Cauchy hypersurfaces and use it in order to show that the
evolution of scalars and vectors fields in the 3rd spatial direction
can be constructed by using a Hamilton-like procedure. Finally, we
discuss the relevance of this result to quantum gravity. 
\end{abstract}

\section*{Introduction}

The correct way in order to obtain a gravitational theory from microscopic
objects or quantum fields is not known but it is expected that it
should be related to surfaces. The main examples are the holographic
principle\cite{holografic}, first proposed by 't Hooft \cite{t Hooft},
which states that in quantum gravity the description of a volume of
space can be encoded on a lower-dimensional boundary to the region,
and the AdS/CFT \cite{ADS/CFT} correspondence, which uses a non-perturbative
formulation of string theory to obtain a realization of the holographic
principle. As far as we know, in all these descriptions the holographic
screen is a light-like surface.

However, recently it seems that non-Cauchy hypersurfaces can also
be related to quantum gravity theories. Non-Cauchy hypersurface foliation
was first used in the membrane paradigm\cite{Membrane} which models
a black hole as a thin, classically radiating membrane, vanishingly
close to the black hole's event horizon. This non-Cauchy hypersurface
is useful for visualizing and calculating the effects predicted by
quantum mechanics for the exterior physics of black holes.

The second example that a non-Cauchy hypersurface is useful for aspects
of quantum gravity involves the surface density of space time degrees
of freedom (DoF). These are expected to be observed by an accelerating
observer in curved spacetime, i.e. whenever an external non-gravitational
force field is introduced \cite{Padmanabhan}. This DoF surface density
was first derived by Padmanabhan for a static spacetime using thermodynamic
considerations. We found that this DoF surface density can also be
constructed from specific canonical conjugate pairs as long as they
are derived in a unique way \cite{merav}. These canonical conjugate
pairs must be obtained by foliating spacetime with respect to the
direction of the external non-gravitational force field. Note that
this aspect reinforces the importance of singling out a very unique
spatial direction: the direction of a non-gravitational force.

The third example which suggests that a non-Cauchy hypersurface is
useful for aspects of quantum gravity involves string theory excitation.
It was found that some specific kind of singularities are obtained
by string theory excitations of a $D1D5$ black hole \cite{D1D5,Mathur:2005zp,Lunin:2002iz,Giusto:2012yz}.
We found \cite{meravLevy} that these singularities can also be explained
using the uncertainty principle between canonical conjugate pairs
which are obtained by singling out the radial direction. The radial
direction can be regarded as the direction of a non- gravitational
force that causes observers to ``stand steel'' in these coordinate
frame. Thus, these singularities, which according to string theory
are expected in quantum gravity theories, are derived by the uncertainty
principle only by singling out the non-gravitational force direction.

The fourth example involves ``holographic quantization”. The holographic
quantization uses spatial foliation in order to quantize the gravitational
fields for different backgrounds in Einstein theory. This is carried
out by singling out one of the spatial directions in a flat background
\cite{Park1} , and also singling out the radial direction for a Schwarzschild
metric \cite{park2}.

The fifth example involves the developing of a quantum black hole
wave packet \cite{Davidson:2014tda}. In this case, the gravitational
foliation is used in order to obtain a quantum Schwarzschild black
hole, at the mini super spacetime level, by a wave packet composed
of plane wave eigenstates.

The six and and final example involves the Wheeler-DeWitt metric probability
wave equation. Recently, in \cite{Ghaffarnejad}, foliation in the
radial direction was used to obtain Wheeler-De Witt metric probability
wave equation on the apparent horizon hypersurface of the Schwarzschild
de Sitter black hole. By solving this equation, the authors found
that a quantized Schwarzschild de Sitter black hole has a nonzero
value for the mass in its ground state. This property of quantum black
holes leads to stable black hole remnants.

All these suggest that in order to find a quantum gravity theory,
one needs to obtain a proper way of constructing a quantum theory
using non-Cauchy hypersurface foliations.

However, in general, one should expect problems when foliating with
respect to hypersurfaces that are not Cauchy, since in this case the
field evolution isn't usually causal. Even constructing a quantum
theory with causal commutation relation on a non-Cauchy hypersurface
is expected to be challenging, since the usual Poisson brackets do
not lead to causal commutation relation on the non-Cauchy hypersurface
and thus are not relevant in this case.

In this paper we propose a method to identify a set of causal canonical
classical brackets on hypersurfaces that are not Cauchy. We use this
method to derive the field equations for scalar fields and find that
the expected field equations can be derived only if the scalar field
is physical on the non-Cauchy hypersurface. This means that as long
as a scalar field is physical on the hypersurface one can use non-Cauchy
foliation to derive causal evolution. Since this can easily be extended
to any vector field, our proof is expected to be relevant even for
the gravitational field in Minkowski spacetime. Thus, when using spatial
foliation, a causal evolution can be constructed for the gravitational
field, as long as the gravitational field is physical on the surface.
However, according to our method, in order to be able to use the spatial
foliation we need to obtain the classical brackets from the $quantum$
commutation relation on the non-Cauchy hypersurface. This is problematic
since we do not have a proper quantum gravitational theory. In this
paper we suggest a method in order overcome this problem, and derive
a possible renormalized quantum gravity theory.

This paper is organized as follows: First we present the problem of
using the Poisson brackets on a non-Cauchy hypersurface. Next we suggest
a method for identifying causal classical brackets between the fields
on the hypersurface. Then we use this method for free scalar field
and find that the Klein-Gordon equation can be derived by a Hamiltonian-like
formalism. Finally, we extend this to any vector field and discuss
the relevance of this result to quantum gravity theories.

\section*{Presenting the problem and the suggested solution}

In general, in order to quantize a field theory we usually start with
the Lagrangian density of the theory $\mathscr{L}(\phi(x),\partial_{\mu}\phi(x))$
and a Cauchy surface which can be defined as an equal time surface
$x_{0}=0$. Then we use to define the momentum canonically conjugate
to the field variable $\phi(x)$ as $\Pi(x)=\frac{\partial\mathscr{L}}{\partial{\dot{\phi}}(x)}$
where $\dot{\phi}=\partial_{0}\phi$, and the Hamiltonian density
as $\mathscr{H}=\Pi(x)\dot{\phi}-\mathscr{L}$. Next, we need to verify
whether the dynamical equation derived by the Euler-Lagrange equations,
can be written in the Hamiltonian form: $\dot{\phi}(x)=\{\phi(x),H\}$
and $\dot{\Pi}(x)=\{\Pi(x),H\}$, where $H=\int d^{3}x\mathscr{H}$.
For this purpose we need to identify equal time canonical bracket
relations between the field variable $\phi(x)$ and the conjugate
momentum $\Pi(y)$. Usually, we assume the equal time canonical Poisson
bracket relations: $\{\phi(x),\phi(y)\}_{x^{0}=y^{0}}=\{\Pi(x),\Pi(y)\}_{x^{0}=y^{0}}=0$
and $\{\phi(x),\Pi(y)\}_{x^{0}=y^{0}}=\delta^{3}(\textbf{x}-\textbf{y})$.
Then, if the fields equations can be written in the Hamiltonian form,
we can treat $\phi(x)$ and $\Pi(x)$ as operators that satisfy the
same equal time commutation relations, i.e.: $\{A(x),B(y)\}_{x^{0}=y^{0}}\rightarrow i\hbar\left[A(x),B(y\right]_{x^{0}=y^{0}}$,
and the quantization is straightforward.

In this paper we want to repeat this process of quantization for a
non-Cauchy hypersurface which we define as a hypersurface with an
equal spatial coordinate $x_{1}=const.$ The other coordinates on
this hypersurface will be denoted by $\tilde{x}=(x^{0},x^{2},x^{3})$.
In this case we propose that the quantization will be as follows:
we will define a new canonically conjugate momentum to the field variable
$\phi(x)$: $\Pi_{1}(x)=\frac{\partial\mathscr{L}}{\partial{\phi'}(x)}$
where $\phi'=\partial_{1}\phi$. The new Hamiltonian-like density
will be defined as $\mathscr{H}_{1}=\Pi_{1}(x)\phi'-\mathscr{L}$.
Next we will need to verify that the dynamical equations derived by
Euler-Lagrange equations could be written in a Hamiltonian-like form
as $\phi(x)'=\{\phi(x),H_{1}\}$ and $\Pi_{1}(x)'=\{\Pi_{1}(x),H_{1}\}$,
where $H_{1}=\int d^{3}\widetilde{x}\mathscr{H_{1}}$. For this purpose
we will need to assume equal $x_{1}$ bracket relations between the
field variable $\phi(x)$ and the new canonical conjugate momentum
$\Pi_{1}(x)$. However, if we will use equal $x_{1}$ canonical Poisson
bracket relations in the same way as equal $x^{0}$ canonical Poisson
bracket relations, namely:$\{\phi(x),\phi(y)\}_{x^{1}=y^{1}}=\{\Pi_{1}(x),\Pi_{1}(y)\}_{x^{1}=y^{1}}=0$
and $\{\phi(x),\Pi_{1}(y)\}_{x^{1}=y^{1}}=\delta^{3}(\widetilde{x}-\widetilde{y})$,
we will find non causal relations. Thus, equal $spatial$ coordinate
brackets can not be defined by the ordinary canonical Poisson brackets.
In order to identify correctly the equal $x_{1}$ classical bracket
relations between the field variable $\phi(x)$ and the new canonically
conjugate momentum $\Pi_{1}(x)$, we will need to have an extension
of the canonical Poisson brackets in such a way that it will be causally
defined even when foliating spacetime with respect to non-Cauchy hypersurfaces.

We suggest a way for identifying the equal $x_{1}$ canonical classical
brackets without using the extended Poisson brackets: It seems that
for any theory that we are able to quantize using the ordinary equal
time canonical Poisson brackets, we can easily obtain the quantum
commutation relation for any equal spatial coordinate. This can be
done by using the field equations and calculating the commutation
relation $\left[\phi(x),\phi(y)\right]$ for any two points $x$ and
$y$ in space-time. This will enable us to calculate the equal spatial
commutation relations $\left[\phi(x),\phi(y)\right]_{x^{1}=y^{1}}$,
$\left[\phi(x),\Pi_{1}(y)\right]_{x^{1}=y^{1}}$ and $\left[\Pi_{1}(x),\Pi_{1}(y)\right]_{x^{1}=y^{1}}$.
Thus from the calculated quantum commutation relations we deduce the
classical equal spatial coordinate brackets between the fields.

\section*{Example: the free scalar field in (1+3)D}

\subsubsection*{1. Spatial foliation and Hamilton-like equation}

We begin with the Lagrangian density of a free scalar field in 1+3
dimension: $\mathscr{L}=\frac{1}{2}\partial_{\mu}\phi\partial^{\mu}\phi-\frac{m^{2}}{2}\phi^{2}$,
and a non-Cauchy equal hypersurface $x_{1}=const$. Then we define
\begin{eqnarray}
\Pi_{1}(x) & =\frac{\partial\mathscr{L}}{\partial{\phi'}(x)} & =-\partial_{1}\phi,\label{new momnta-1}
\end{eqnarray}
and the new Hamiltonian-like density becomes: 
\begin{eqnarray}
\mathscr{H}_{1} & = & \Pi_{1}(x)\phi'-\mathscr{L}=\nonumber \\
 & = & \frac{1}{2}\left(-\Pi_{1}^{2}-(\partial_{0}\phi)^{2}+(\partial_{2}\phi)^{2}+(\partial_{3}\phi)^{2}+m^{2}\phi^{2}\right).\label{eq:hamiltonian}
\end{eqnarray}
Then the new Hamiltonian-like equations are: 
\begin{eqnarray}
\phi(x)' & = & \{\phi(x),H_{1}\}\label{hamilton like-1}\\
\Pi_{1}(x)' & = & \{\Pi_{1}(x),H_{1}\}\label{eq:4}
\end{eqnarray}
where $H_{1}=\int d^{3}\widetilde{y}\mathscr{H}_{1}(\tilde{y,}y_{1})$.

Note that, since $H_{1}$ is $y_{1}$ independent\footnote{This can easily be proven using the Hamilton formalism where we expect
that for any $A(x)$ : $A(x)'=\left\{ A(x),H_{1}(x^{1})\right\} $
thus for $A(x)=\mathscr{H}_{1}(x)$ we find $\mathscr{H}_{1}(x)'=\left\{ \mathscr{H}_{1}(x),H_{1}(x^{1})\right\} $.
Using $H_{1}(x^{1})=\int dx^{0}dx^{2}dx^{3}\mathscr{H}_{1}$ we find
$H_{1}'(x^{1})=\left\{ H_{1}(x^{1}),H_{1}(x^{1})\right\} =0$}, we can choose $y_{1}=x_{1}$.

Obviously, when going to the quantum limit attention is needed in
order to construct the new Hamiltonian and the apparent phase space
on a time-like surface. For example, it is problematic to quantize
such Hamiltonians since they are unbounded from below. Even operator
ordering ambiguities are expected and must be redefined. Moreover,
we used to expect that before going to the quantum limit one requires
a geometric symplectic structure which will lead to the definition
of the phase space and the conjugate relations. However, as we will
demonstrate in the next section, this symplectic structure is not
relevant on time-like surfaces. 

\subsubsection*{2. Finding the equal $x_{1}$ canonical classical brackets for a
free scalar field}

In order to identity the equal $x_{1}$ classical bracket relations,
we use the commutation relation $\left[\phi(x),\phi(y)\right]$, derived
by the ordinary quantization \cite{das}, for any two points $x$
and $y$ in space-time:

\begin{eqnarray}
\left[\phi(x),\phi(y)\right] & = & -i\int\frac{d^{3}\textbf{k}}{(2\pi)^{3}}\frac{sin[E_{\textbf{k}}(x^{0}-y^{0})]}{E_{\textbf{k}}}e^{i\textbf{k}(\textbf{x}-\textbf{y})}\label{commutation relation-1}
\end{eqnarray}
where $E_{\textbf{k}}=\sqrt{\textbf{k}^{2}+m^{2}}=\sqrt{k_{1}^{2}+k_{2}^{2}+k_{3}^{2}+m^{2}}$.

It is useful to write this commutation relation in a covariant way,
using the Heaviside step function $\theta(x)$: 
\begin{eqnarray}
\left[\phi(x),\phi(y)\right] & = & \int\frac{d^{4}k}{(2\pi)^{3}}\left(\theta(k^{0})-\theta(-k^{0})\right)\delta(k^{2}-m^{2})e^{ik(x-y)}.\label{commutation relation1-1}
\end{eqnarray}
Next, using $\delta(k^{2}-m^{2})=\delta\left((k^{1})^{2}-P_{x}^{2}\right)=\frac{1}{2P_{x}}\left[\delta(k^{1}-P_{x})+\delta(k^{1}+P_{x})\right]$
where $P_{x}\equiv\sqrt{(k^{0})^{2}-(k^{2})^{2}-(k^{3})^{2}-m^{2}}$
, we find that the same commutation relations equal also to: 
\begin{eqnarray}
\left[\phi(x),\phi(y)\right] & = & \int\frac{d^{4}k}{(2\pi)^{3}}\left(\theta(k^{0})-\theta(-k^{0})\right)\frac{1}{2P_{x}}\left[\delta(k^{1}-P_{x})+\delta(k^{1}+P_{x})\right]e^{ik(x-y)}\nonumber \\
 & = & \int\frac{d^{3}\widetilde{k}}{(2\pi)^{3}}\left(\theta(k^{0})-\theta(-k^{0})\right)\theta\left(P_{x}^{2}\right)\frac{e^{iP_{x}(x^{1}-y^{1})}+e^{-iP_{x}(x^{1}-y^{1})}}{2P_{x}}e^{i\widetilde{k}\cdot(\widetilde{x}-\widetilde{y})}\nonumber \\
 & = & \int\frac{d^{3}\widetilde{k}}{(2\pi)^{3}}\epsilon(\pm k^{0},P_{x}^{2})\frac{cos[P_{x}(x^{1}-y^{1})]}{P_{x}}e^{i\widetilde{k}\cdot(\widetilde{x}-\widetilde{y})},\label{commutation relation2-1}\\
\nonumber 
\end{eqnarray}
where $\widetilde{k}=(k^{0},k^{2},k^{3})$ , $\widetilde{k}\cdot(\widetilde{x}-\widetilde{y})=k^{0}(x^{0}-y^{0})-k^{2}(x^{2}-y^{2})-k^{3}(x^{3}-y^{3})$,
and $\epsilon(\pm k^{0},P_{x}^{2})\equiv\left(\theta(k^{0})-\theta(-k^{0})\right)\theta\left(P_{x}^{2}\right)$.
Note that the term $\theta\left(P_{x}^{2}\right)$ actually limits
the possible values of $\widetilde{k}$ so that only physical modes
are considered. The fact that we are only considering physical modes
is important in our derivation of the field equations. From (\ref{commutation relation2-1})
we see that the equal spatial coordinate $(x^{1}=y^{1})$ commutation
relations become:

\begin{equation}
\left[\phi(x),\phi(y)\right]_{x^{1}=y^{1}}=\int\frac{d^{3}\widetilde{k}}{(2\pi)^{3}}\epsilon(\pm k^{0},P_{x}^{2})\frac{1}{P_{x}}e^{i\widetilde{k}\cdot(\widetilde{x}-\widetilde{y})}.
\end{equation}
Note that this commutation relations are indeed causal since they
do not vanish for $x$ and $y$ which are causally connected.

Next we need to evaluate the equal spatial coordinate $(x^{1}=y^{1})$
commutation relations between $\phi(x)$ and its new canonical conjugate
momentum $\Pi_{1}(y)$. Varying eq. (\ref{commutation relation2-1})
with respect to $y^{1}$ and using: $\Pi_{1}(y)=-\frac{\partial\phi(y)}{\partial y_{1}}$
gives: 
\begin{equation}
\left[\phi(x),\Pi_{1}(y)\right]=-\int\frac{d^{3}\widetilde{k}}{(2\pi)^{3}}\epsilon(\pm k^{0},P_{x}^{2})sin[P_{x}(x^{1}-y^{1})]e^{i\widetilde{k}\cdot(\widetilde{x}-\widetilde{y})}\label{eq:pai phy relation}
\end{equation}
and the equal spatial coordinate $x^{1}=y^{1}$ commutation relations
become: 
\begin{eqnarray}
\left[\phi(x),\Pi_{1}(y)\right]_{x^{1}=y^{1}} & = & 0.\label{commutation relation5-1}
\end{eqnarray}
Thus we got that for a free scalar field $\phi(x)$ and $\Pi_{1}(y)$
commutes on $x^{1}=y^{1}$ and do not influence each other even when
their coordinates are causally connected.

Finally, we evaluate the equal spatial coordinate $(x^{1}=y^{1})$
commutation relations between the two canonical conjugate momenta
$\Pi_{1}(x)$ and $\Pi_{1}(y)$ . Varying eq. (\ref{eq:pai phy relation})
with respect to $x^{1}$ once again and using: $\Pi_{1}(x)=-\frac{\partial\phi(x)}{\partial x_{1}}$
gives: 
\begin{equation}
\left[\Pi_{1}(x),\Pi_{1}(y)\right]=\int\frac{d^{3}\widetilde{k}}{(2\pi)^{3}}\epsilon(\pm k^{0},P_{x}^{2})P_{x}cos[P_{x}(x^{1}-y^{1})]e^{i\widetilde{k}\cdot(\widetilde{x}-\widetilde{y})}
\end{equation}
and thus the equal spatial coordinate $(x^{1}=y^{1})$ commutation
relations between the $\Pi_{1}$ fields are: 
\begin{eqnarray}
\left[\Pi_{1}(x),\Pi_{1}(y)\right]_{x^{1}=y^{1}}=\int\frac{d^{3}\widetilde{k}}{(2\pi)^{3}}\epsilon(\pm k^{0},P_{x}^{2})P_{x}e^{i\widetilde{k}\cdot(\widetilde{x}-\widetilde{y})}.\label{commutation relation7-1}
\end{eqnarray}

Now we take a very naive assumption and suggest what the $classical$
\textquotedbl{}Poisson-like\textquotedbl{} brackets for an equal spatial
coordinate can be achieved just by multiplying the commutation relation
we have got by $-i$. We have: 
\begin{eqnarray}
\{\phi(x),\phi(y)\}_{x^{1}=y^{1}} & = & -i\int\frac{d^{3}\widetilde{k}}{(2\pi)^{3}}\epsilon(\pm k^{0},P_{x}^{2})\frac{1}{P_{x}}e^{i\widetilde{k}\cdot(\widetilde{x}-\widetilde{y})}\label{commutation relation8-1}\\
\{\Pi_{1}(x),\Pi_{1}(y)\}_{x^{1}=y^{1}} & = & -i\int\frac{d^{3}\widetilde{k}}{(2\pi)^{3}}\epsilon(\pm k^{0},P_{x}^{2})P_{x}e^{i\widetilde{k}\cdot(\widetilde{x}-\widetilde{y})}\label{eq:14}\\
\{\phi(x),\Pi_{1}(y)\}_{x^{1}=y^{1}} & = & 0.\label{15}
\end{eqnarray}

As we mentioned before, we used to expect that in order to obtain
the commutation relation in the quantum limit one requires a geometric
symplectic structure of the phase space. The symplectic structure
leads to the definition of phase space, conjugate relations and their
evolution. However, as we see from eqs. (\ref{commutation relation8-1},\ref{eq:14},\ref{15})
the symplectic structure is not relevant on time-like surfaces. Thus
though it is not clear whether one can construct the phase space in
this case, we suggest examining whether these seemingly canonical
fields can lead to the expected evolution, i.e. the field equation.

\subsubsection*{3. The dynamical equation derived by the Hamiltonian-like equations}

Finally, we must check whether the dynamical equation derived by the
Euler-Lagrange equation; i.e. the Klein-Gordon equation: $(\partial_{\mu}\partial^{\mu}+m^{2})\phi=0$
can be constructed from the Hamiltonian-like equations (\ref{hamilton like-1})
and(\ref{eq:4}) when the classical brackets (\ref{commutation relation8-1})-(\ref{15})
are assumed.

.

We start with (\ref{hamilton like-1}), and use (\ref{eq:hamiltonian})
and get

\[
\phi(x)'=\left\{ \phi(x),\int d^{3}\widetilde{y}\left(\frac{1}{2}\left(-\Pi_{1}(y)^{2}-(\partial_{0}\phi(y))^{2}+(\partial_{2}\phi(y))^{2}+(\partial_{3}\phi(y))^{2}+m^{2}\phi(y)^{2}\right)\right)_{y_{1}=x_{1}}\right\} .
\]
Using the vanishing equal $x_{1}$bracket relation $\phi$ and $\Pi_{1}$
we have:

\begin{equation}
\phi(x)'=\int d^{3}\widetilde{y}\left\{ \phi(x),\phi(y)\right\} _{x^{1}=y^{1}}\left(\partial_{0}^{2}\phi(y)-\partial_{2}^{2}\phi(y)-\partial_{3}^{2}\phi(y)+m^{2}\phi(y)\right){}_{y_{1}=x_{1}},\label{eq:phy tag}
\end{equation}
and after using the equal $x_{1}$ bracket relations (\ref{commutation relation8-1})
we get after integration by parts:

\begin{equation}
\phi(x)'=i\int d^{3}\widetilde{y}\int\frac{d^{3}\widetilde{k}}{(2\pi)^{3}}\epsilon(\pm k^{0},P_{x}^{2})P_{x}e^{i\widetilde{k}\cdot(\widetilde{x}-\widetilde{y})}\phi(y)_{x^{1}=y^{1}}.\label{eq:phi tag 2-1}
\end{equation}
Varying (\ref{eq:phi tag 2-1}) with respect to $x_{1}$ once again
we get:

\begin{equation}
\phi(x)''=i\int d^{3}\widetilde{y}\int\frac{d^{3}\widetilde{k}}{(2\pi)^{3}}\epsilon(\pm k^{0},P_{x}^{2})P_{x}e^{i\widetilde{k}\cdot(\widetilde{x}-\widetilde{y})}\phi(y)'_{x^{1}=y^{1}}.
\end{equation}
and using (\ref{eq:phi tag 2-1}) for $\phi(y)'$ we find: 
\begin{equation}
\phi(x)''=-\int d^{3}\widetilde{y}'\phi(y')_{x^{1}=y'^{1}}\int\frac{d^{3}\widetilde{k}}{(2\pi)^{3}}\epsilon^{2}(\pm k^{0},P_{x}^{2})P_{x}^{2}\left(e^{i\widetilde{k}\cdot(\widetilde{x}-\widetilde{y}')}\right)
\end{equation}

Since \footnote{$\theta^{2}(k^{0})=\theta(k^{0})$ and $\theta(k^{0})\theta(-k^{0})=0$}
$\epsilon^{2}(\pm k^{0},P_{x}^{2})$$=\theta(P_{x}^{2})$ we finally
have: 
\begin{equation}
\phi(x)''=-\int\frac{d^{3}\widetilde{k}}{(2\pi)^{3}}\theta(P_{x}^{2})\left((k^{0})^{2}-(k^{2})^{2}-(k^{3})^{2}-m^{2}\right)\int d^{3}\widetilde{y}'e^{i\widetilde{k}\cdot(\widetilde{x}-\widetilde{y}')}\phi(y')_{x^{1}=y'^{1}}.\label{eq:phi tagiim}
\end{equation}
Now, Fourier transforming this equation with respect to $\tilde{k}$
yields:

\begin{equation}
\left(k^{1}\right)^{2}=\left[(k^{0})^{2}-(k^{2})^{2}-(k^{3})^{2}-m^{2}\right]\cdot\theta\left[(k^{0})^{2}-(k^{2})^{2}-(k^{3})^{2}-m^{2}\right].\label{eq:levy2}
\end{equation}
Thus, if we limit ourselves to physical fields i.e.:

\[
(k^{0})^{2}-(k^{2})^{2}-(k^{3})^{2}-m^{2}\geq0,
\]
we get: 
\begin{equation}
(k^{0})^{2}-(k^{1})^{2}-(k^{2})^{2}-(k^{3})^{2}-m^{2}=0\label{eq:levy}
\end{equation}
which are the expected fields equation.

.

To conclude: we found that quantizing a scalar field by foliating
spacetime with respect to a spacelike vector is possible.

.

The same procedure can be used in order to find an equation for the
$\Pi_{1}$ field. Using (\ref{eq:4}) and (\ref{eq:hamiltonian})
we have:

\[
\Pi_{1}(x)'=\left\{ \Pi_{1}(x),\int d^{3}\widetilde{y}\left(\frac{1}{2}\left(-\Pi_{1}(y)^{2}-(\partial_{0}\phi(y))^{2}+(\partial_{2}\phi(y))^{2}+(\partial_{3}\phi(y))^{2}+m^{2}\phi(y)^{2}\right)\right)_{y_{1}=x_{1}}\right\} .
\]
With (\ref{15}) we get:

\[
\Pi_{1}(x)'=-\int d^{3}\widetilde{y}\left\{ \Pi_{1}(x),\Pi_{1}(y)\right\} _{x^{1}=y^{1}}\Pi_{1}(y){}_{y_{1}=x_{1}}
\]
i.e.:

\begin{equation}
\Pi_{1}(x)'=i\int d^{3}\widetilde{y}\int\frac{d^{3}\widetilde{k}}{(2\pi)^{3}}\epsilon(\pm k^{0},P_{x}^{2})P_{x}e^{i\widetilde{k}\cdot(\widetilde{x}-\widetilde{y})}\Pi_{1}(y){}_{y_{1}=x_{1}}\label{eq:pay tag-1}
\end{equation}
which is the same equation which the field $\phi$ fulfills.

Thus, though the symplectic structure is not relevant on time-like
surfaces, we got two Hamilton-like independent field equations for
the fields $\phi$ and $\Pi_{1}$.

This result can easily extended for any vector fields. For vector
fields, one can ignore complication associated with gauge invariance
and work directly with physical components. In this case the action
of each physical component will be the same as for a scalar field.
For example, though in $(3+1)D$ the metric has 10 components 8 of
them are non-physical, and each of the two remaining physical components
has an effective action of a scalar field. Thus, quantizing a vector
field by foliating spacetime with respect to a spacelike vector field
is also possible whenever the vector field components have causal
commutation relation on the non-Cauchy hypersurface.

However, the new Hamiltonian and phase space need much more attention
in order to be upgraded to these commutation relations from first
principles on any time-like surface. The encouragement to overcome
the expected difficulties comes from the relevance of such foliations
to quantum gravity theories.

\section*{Implication to quantum gravity theories}

In order to find the way to a renormalized quantum gravity, we need
to derive the true degrees of freedom of quantum gravity. Various
reductions of the quantum gravity degrees of freedom were reported
in the past in \cite{past York,past Fischer,past Gay-Balmaz,past Gomes}.
All these works employed the usual $(3+1)D$ splitting with the genuine
time coordinate separated out. However, in light of the holographic
picture, it is worthwhile considering the possible advantages of a
non-Cauchy surface foliation on quantum gravity theories and its expected
benefit to the renormalization problem of these theories. Park \cite{Park1}
separates the spatial directions. His strategy for reduction has been
the removal of all of the nonphysical degrees of freedom from the
external states. His key observation for the reduction was the fact
that the residual 3D gauge symmetry - whose detailed analysis is given
in \cite{park4} - can be employed to gauge away the non-dynamical
fields. However, this kind of reduction is not always possible.

We suggest that, whenever it is possible to obtain a renormalized
$(2+1)D$ quantum gravitational theory only on a specific non-Cauchy
hypersurface, we may use this non-Cauchy spatial foliation in order
to obtain a $(3+1)D$ quantum gravitational theory. In order to examine
this option we will first find the conditions in order to obtain a
renormalized $(2+1)D$ quantum gravitational theory only on a specific
non-Cauchy hypersurface. Next we will suggest a way to construct a
causal $(3+1)D$ quantum gravitational theory. Finally, we will discus
the implication of this procedure on the expected properties of quantum
gravity theories and on the renormalization possibility of a $(3+1)D$
quantum gravity theory.

Lets start by considering the standard foliation of spacetime with
respect to some spacelike hypersurfaces whose directions are $n^{a}$.
The lapse function $M$ and shift vector $W_{a}$ satisfy $r_{a}=Mn_{a}+W_{a}$
where $r^{a}\nabla_{a}r=1$ and $r$ is constant on $\Sigma_{\text{r}}$.
The $\Sigma_{r}$ hyper-surfaces metric $h_{ab}$ is given by $g_{ab}=h_{ab}+n_{a}n_{b}$.
The extrinsic curvature tensor of the hyper-surfaces is given by $K_{ab}=-\frac{1}{2}\mathcal{L}_{n}h_{ab}$
where $\mathcal{L}_{n}$ is the Lie derivative along $n^{a}$. \footnote{The intrinsic curvature $R_{ab}^{(3)}$ is then given by the 2+1 Christoffel
symbols: $\Gamma_{ab}^{k}=\frac{1}{2}h^{kl}\left(\frac{\partial h_{lb}}{\partial x^{a}}+\frac{\partial h_{al}}{\partial h^{b}}-\frac{\partial h_{ab}}{\partial x^{l}}\right)$
so that $R_{ab}^{(3)}=\frac{\partial\Gamma_{ab}^{k}}{\partial x^{k}}-\frac{\partial\Gamma_{ak}^{k}}{\partial x^{b}}+\Gamma_{ab}^{k}\Gamma_{kl}^{l}-\Gamma_{al}^{l}\Gamma_{lb}^{k}$
.} Instead of the$(3+1)D$ Einstein equations

\[
R_{ab}^{(4)}=8\pi\left(T_{ab}-\frac{1}{2}Tg_{ab}\right)
\]
one finds \cite{book 3+1}: 
\[
R_{ab}^{(3)}+KK_{ab}-2K_{ai}K_{b}^{i}-M^{-1}\left(\mathcal{L}_{r}K_{ab}+D_{a}D_{b}M\right)=8\pi\left(S_{ab}-\frac{1}{2}\left(S-P\right)h_{ab}\right),
\]
\[
R^{(3)}+K^{2}-K_{ab}K^{ab}=16\pi P,
\]
\[
D_{b}K_{a}^{b}-D_{a}K=8\pi F_{a}.
\]
Where $D_{a}$ represent the 2+1 covariant derivatives, $S_{ab}=h_{ac}h_{ad}T^{cd}$
, $P=n_{c}n_{d}T^{cd}$ and $F_{a}=h_{ac}n_{b}T^{cb}$.

Lets assume that on some unique hypersurface $r=r_{0}$:

\[
KK_{ab}-2K_{ai}K_{b}^{i}-M^{-1}\left(\mathcal{L}_{r}K_{ab}+D_{a}D_{b}M\right)=0
\]
then we have: 
\[
R_{ab}^{(3)}=8\pi\left(S_{ab}-\frac{1}{2}\left(S-P\right)h_{ab}\right),
\]
which is just the Einstein equation in 2+1 dimension, when $P$ serves
as a cosmological constant.

From the quantum gravity point of view, this unique hypersurface is
interesting. To see this note that though we don't know how to obtain
a renormalized quantum gravity theory in 3+1 dimensions, a renormalized
quantum theory in 2+1 dimensions is possible \cite{carlip}. Thus,
when the $(3+1)D$ Einstein equations reduce to a kind of 2+1 Einstein
equations on some hypersurface $r=r_{0}$, we can quantize the gravitational
fields on the hypersurface $r=r_{0}$ at least with respect to this
foliation.

Now, as we have got this$(2+1)D$ Einstein equation on this spacelike
non-Cauchy hypersurface, lets assume we have found a $(2+1)D$ renormalized
quantum gravitational theory on the hypersurface $r=r_{0}$ . As we
noted before, having the quantum theory on the non-Cauchy hypersurface
means one can find also a set of causal commutation relation on this
hypersurface. But, as we showed in the last section by an example,
knowing the causal commutation relation on a non-Cauchy hypersurface
automatically enables us to know how the quantum gravity fields will
evolve through the remaining spatial direction $r\neq r_{0}$, with
the aid of the new Hamiltonian-like equations we mentioned. Thus,
using a sort of Hamilton-like ADM equation may serve as a way of finding
the evolution of the gravitational quantum fields in the third spatial
direction and thus of producing, at least in principle, a$(3+1)D$
renormalized quantum gravity theory.

Note that this procedure is related to holography. To see this note
that in this formalism, the evolution on the gravitational fields
in the $4-th$ dimension is determined by the \textquotedbl{}initial\textquotedbl{}
condition on the non-Cauchy hyper-surface $r=r_{0}$. Thus in this
formalism all the information needed to describe the evolution of
the gravitational field in the $4-th$ dimension in encoded on this
hyper-surface, as suggested by holography.

\section*{Summary}

In this paper we argued that deriving a proper quantum gravity theory
may involve quantization on a non-Cauchy hypersurface. We showed that
since in this kind of hypersurfaces the ordinary Poisson brackets
are not causal, constructing a quantum theory on a non-Cauchy hypersurface
is expected to be problematic. We suggested a method to identify classical
brackets that are causal even on a non-Cauchy hypersurface. We used
this method to identify the (causal) brackets between free relativistic
scalar fields and their (new) canonical conjugate momentum on a non-Cauchy
hypersurface. Next we proposed a sort of classical Hamilton-like equations
for the evolution along the direction perpendicular to the non-Cauchy
hypersurface. We found that, as long as the field on the non-Cauchy
hypersurface is physical, these equations leads to the expected free
Klein-Gordon equation. 

However, the new Hamiltonian and phase space need more attention in
order to be upgraded to the commutation relations we obtained from
first principles on any time-like surface. To begin with, the probability
definition can only be defined on a Cauchy surface \footnote{To see this note that any attempt to derive the state normalizability
on a time-like surface leads to temporal decay, both in past and future,
making the usual norm of a state a time-dependent object. }. Thus, in this case, it seems that in order to obtain the probability
density one needs to consider unitarity as a constraint. Moreover,
the new Hamiltonians seem to be unbounded from below and operator
ordering ambiguities are expected and must be redefined. Although
this is only a partial list of the expected ambiguities when quantum
theory is derived by spatial foliation, the expected properties of
a quantum gravity theory lead us to suggest that such a theory may
be constructed using spatial foliation.

Thus, we considered the possible advantages of such unique foliation
on expected theories of quantum gravity. We considered a foliation
of spacetime with respect to spacelike hypersurfaces whose directions
are $n^{a}$, and the conditions for the $(3+1)D$ Einstein equations
to reduce to a kind of $(2+1)D$ Einstein equations on a hypersurface
$r=r_{0}$. Then, by using the argument that a renormalized quantum
gravity theory can be constructed in $(2+1)D$, we discussed the procedure
needed in order to derive the renormalized gravitational fields causal
evolution in the 3rd spatial direction. However, whether this suggested
procedure is expected to leads to a renormalized quantum gravity theory
on $r\neq r_{0}$ remains to be seen.

\textbf{Acknowledgments:} We thank Oded Kenneth for reading the first
manuscript of the paper and for his fruitful comments. This research
was supported by The Open University of Israel's Research Fund (grant
no. 509565).

\end{document}